\documentclass[
11pt,final,notitlepage,oneside,twocolumn,nobibnotes,nofootinbib,superscriptaddress,noshowpacs,centertags]
{revtex4}

\RequirePackage{amssymb,amsmath}
\RequirePackage{eufrak}
\RequirePackage{mathtext}
\RequirePackage[koi8-r]{inputenc}
\RequirePackage[T2A]{fontenc}
\RequirePackage[russian,english]{babel}
\RequirePackage{graphicx}
\RequirePackage{bm}
\RequirePackage{array}
\RequirePackage{longtable}
\RequirePackage{graphicx}
\RequirePackage{calc}
\RequirePackage{ifthen}

\graphicspath{{../images/}}

\def\saoname{Special Astrophysical Observatory of the Russian AS, Nizhnij Arkhyz 369167, Russia}
\def\fm{\hbox{$.\!\!^{\rm m}$}}
\def\sun{\hbox{$\odot$}}
\def\arcsec{\hbox{$^{\prime\prime}$}}
\def\farcs{\hbox{$.\!\!^{\prime\prime}$}}
\newcommand{\ab}{Astrophysical Bulletin}
\newcommand{\aj}{Astronom. J.}
\newcommand{\pasp}{Publ. Astronom. Soc. Pacific}
\newcommand{\aaa}{Astronom. and Astrophys.}
\newcommand{\mnras}{Monthly Notices Roy. Astronom. Soc.}
\newcommand{\apjs}{Astrophys. J. Suppl.}


\newcommand{\Ha}{$H\alpha$}
\newcommand{\Hb}{$H\beta$}

\newcommand{\aap}{\aaa}

\newcommand{\aaps}{\apjs}
\newcommand\kms{{\rm\,\mbox{km}\,\mbox{s}$^{-1}$}}
\newcommand{\micron}{\ensuremath{\mu{}m}}

\begin{document}
\title[Two New LBV Candidates in the M\,33 Galaxy]{Two New LBV Candidates in the M\,33 Galaxy}

\author{A.~F.~Valeev}
\affiliation{\saoname}

\author{{O.~N.}~{Sholukhova}}
\affiliation{\saoname}

\author{{S.~N.}~{Fabrika}}
\affiliation{\saoname}

\begin{abstract}
We present two new luminous blue variable (LBV) candidate stars
discovered in the M\,33 galaxy. We identified these stars (Valeev
et al. \cite{ValeevCatalogLBV2010}) as massive star candidates at
the final stages of evolution, presumably with a notable
interstellar extinction. The candidates were selected from the
Massey et al. catalog \cite{Massey2006UBVRIcatalog} based on the
following criteria: emission in \Ha,  V<18\fm5 and $0\fm35 < (B-V)
< 1\fm2$. The spectra of both stars reveal a broad and strong
\Ha{} emission with extended wings (770 and 1000  \kms{}). Based
on the spectra we estimated the main parameters of the stars.
Object N\,45901 has a bolometric luminosity
\mbox{$\log$(L/L$_{\sun}$) = 6.0 -- 6.2} with the value of
interstellar extinction A$_V = 2.3 \pm 0.1$. The temperature of
the star's photosphere is estimated as \mbox{T$_\star \sim 13000
- 15000$\,K,} its probable mass on the Zero Age Main Sequence is
\mbox{M $\sim$ 60--80~M$_{\sun}$.} The infrared excess in N
\,45901 corresponds to the emission of warm dust with the
temperature T$_{\mbox{warm}} \sim 1000$\,K, and amounts to 0.1\,\%
of the bolometric luminosity. A comparison of stellar magnitude
estimates from different catalogs points to the probable
variability of the object N\,45901. Bolometric luminosity of the
second object, N\,125093, is $\log$(L/L$_{\sun}$) = 6.3 -- 6.6,
the value of interstellar extinction is \mbox{A$_V = 2.75 \pm
0.15$.} We estimate its photosphere's temperature as T$_\star \sim
13000 - 16000$\,K, the initial mass as M $\sim$ 90-120~M$\sun$.
The infrared excess in N\,125093 amounts to 5--6\,\% of the
bolometric luminosity. Its spectral energy distribution reveals
two thermal components with the temperatures
\mbox{T$_{\mbox{warm}} \sim 1000$\,K} and \mbox{T$_{\mbox{cold}}
\sim 480$\,K.} The \mbox{[Ca\,II]\,$\lambda \lambda 7291, 7323$}
lines, observed in LBV-like stars Var\,A and N\,93351 in M\,33 are
also present in the spectrum of N\,125093. These lines indicate
relatively recent gas eruptions and dust activity linked with
them. High bolometric luminosity of these stars and broad \Ha{}
emissions allow classifying the studied objects as LBV candidates.
\end{abstract}
\email{azamat@sao.ru} 

\maketitle

\section{INTRODUCTION}

Luminous blue variables are the most massive stars at one of the
final stages of stellar evolution \cite{HumphreysDavidson1994}. A
small number of massive stars in the galaxies and a very short
scale of the LBV phase   ($10^4-10^5$ years) makes these objects
unique. The study of LBV-type stars is essential for understanding
the evolution of stars on the upper part of the Main Sequence, the
formation of WR stars, supernovae, relativistic stars (black
holes), for the comprehension of the mechanisms of mass loss, and
heavy-element enrichment in galaxies.

It is believed that the stars with more than 40\,M$_{\sun}$
undergo the LBV phase  \cite{Meynet2007}. However, the
relationship between the LBV stars and other types of massive
stars at the final stages of evolution (red supergiants RSG,
yellow supergiants YSG, blue supergiants BSG, B[e]-supergiants, WR
stars), and evolutionary transitions between these types of stars
are still unclear \cite{SmithConti2008}. LBV stars are closely
linked with the late WR stars of the nitrogen sequence with
hydrogen in their atmospheres (WNLh). Two cases of transition are
known from observations: LBV $\to$ WNLh (star V\,532 in M\,33,
\cite{Fabrika2005}) and WN3 $\to$ WN11(LBV) $\to$ WN4/5 (star
HD5980 in the SMC galaxy, \cite{Koenigsberger2010}).

Observational manifestations of LBV stars are very diverse
\cite{vanGenderen2001}, and the number of known objects of this
type is little. It is hence not clear whether the LBV class of
objects is homogeneous. For example, why have the S\,Dor variables
\cite{HumphreysDavidson1994,vanGenderen2001} not revealed giant
eruptions, similar to the  $\eta$ Car and P\,Cyg stars.
It is also difficult to draw specific conclusions about the place
of the LBV phase in the evolution of massive stars, and even more
so about the effects of binarity and rapid rotation at this stage
of evolution.

It is obvious that the above problems can only be solved when a
sufficiently large number of objects is thoroughly studied.
Observations of more LBV-like stars increases our chances to
witness the rare \mbox{LBV $\leftrightarrow$ WN-type} transitions.
This will allow testing the modern evolutionary sequences and the
dependence of various stages of massive star evolution on galactic
metallicity (Z).

Apparently, most of the LBV stars without notable interstellar
extinction in our Galaxy are already discovered. However, it is
probably still possible to discover a few dozen more such objects
in the Galaxy using modern infrared \mbox{surveys
\cite{Gvaramadze2010SpitzerSearch,Gvaramadze2010LBVcand}.} In the
M\,33 galaxy almost all the LBV-like objects  may be detected, as
its fortunate orientation (its inclination to the line of sight is
56 \mbox{deg. \cite{Zaritsky1989M33Inclination}}), and a
relatively close distance \mbox{(950 Mpc,
\cite{Bananos2006M33Distance})} allow detailed spectroscopy of its
bright  stars.

The review \cite{HumphreysDavidson1994} attributes  five and 15
stars to the confirmed members of the LBV class in our Galaxy and
in the galaxies of the Local Group, respectively. In particular,
in the M\,33 galaxy the authors identified four LBV stars.  Clark
et al. \cite{Clark2005} mention 12 LBV stars and 23 LBV candidates
in our Galaxy, while in the M\,33 galaxy 37 LBV candidates are
known \mbox{to date \cite{Massey2007LBVcatalog}.}

According to the catalog \cite{Massey2006UBVRIcatalog}, M\,33
contains 2304 stellar objects with V<18\fm5. If we assume that the
mean interstellar extinction of the brightest stars in the galaxy
is A$_V\approx$1\fm0 (see, e.g., \cite{Fabrika1999}, where this
value is estimated as A$_V\approx$0\fm95$\pm$0\fm05), then with
the distance modulus to M\,33 of \mbox{(m$-$M)$_0$=24\fm9
\cite{Bananos2006M33Distance},} the stars with V<18\fm5 and
$(B-V)$<0\fm35 will \linebreak have the luminosity  M$_V<-$7\fm4
and color \linebreak \mbox{$(B-V)_0\le$ 0\fm0.} In
\cite{ValeevCatalogLBV2010} we made photometry in the ~\Ha~ images
of all the stars from the catalog \cite{Massey2006UBVRIcatalog}
with the above restrictions on color and luminosity, and made a
list of stars with an excess in \Ha{}. These are bright
supergiants of the Iab luminosity class and brighter, and the
hottest Ib supergiants (with B0 spectra and earlier)
\cite{Straizhis}. We expected that all the potential LBV
candidates will make it into our list. We hence isolated in
\cite{ValeevCatalogLBV2010} 185 blue emission objects (V<18\fm5
and \mbox{$(B-V)$<0\fm35),} candidates for massive stars at the
final stages of evolution.

Evidently, LBVs and similar objects may well have the extinction
A$_V$ > 1\fm0, for which reason an {\it additional list} of stars
with V<18\fm5 and \linebreak \mbox{$0\fm35 < (B-V) < 1\fm2$}  with
emission in \Ha~ was made in \cite{ValeevCatalogLBV2010},
containing 25 candidates.

We currently make spectroscopic observations of objects from both
our lists \cite{ValeevCatalogLBV2010}. In the first list we
discovered an LBV-type star N\,93351, which is located in the
nuclear region of M\,33 \cite{Valeev2009NewLBV}. Two objects from
our supplementary list (presumably with a notable own reddening)
were earlier \mbox{classified \cite{Massey2007LBVcatalog}} as hot
LBV candidates: N\,6862 with $B-V =$ 0\fm52 and N\,141751 with
$B-V =$ 0\fm77.

We have recently obtained spectra of 15 candidate stars from the
supplementary list, among them we discovered two new LBV
candidates. In this paper we present the results of study of these
two objects.

\section{OBSERVATIONS AND DATA REDUCTION}
\subsection{Spectral Data}

\begin{figure}[ht]
\includegraphics[width=0.95\columnwidth]{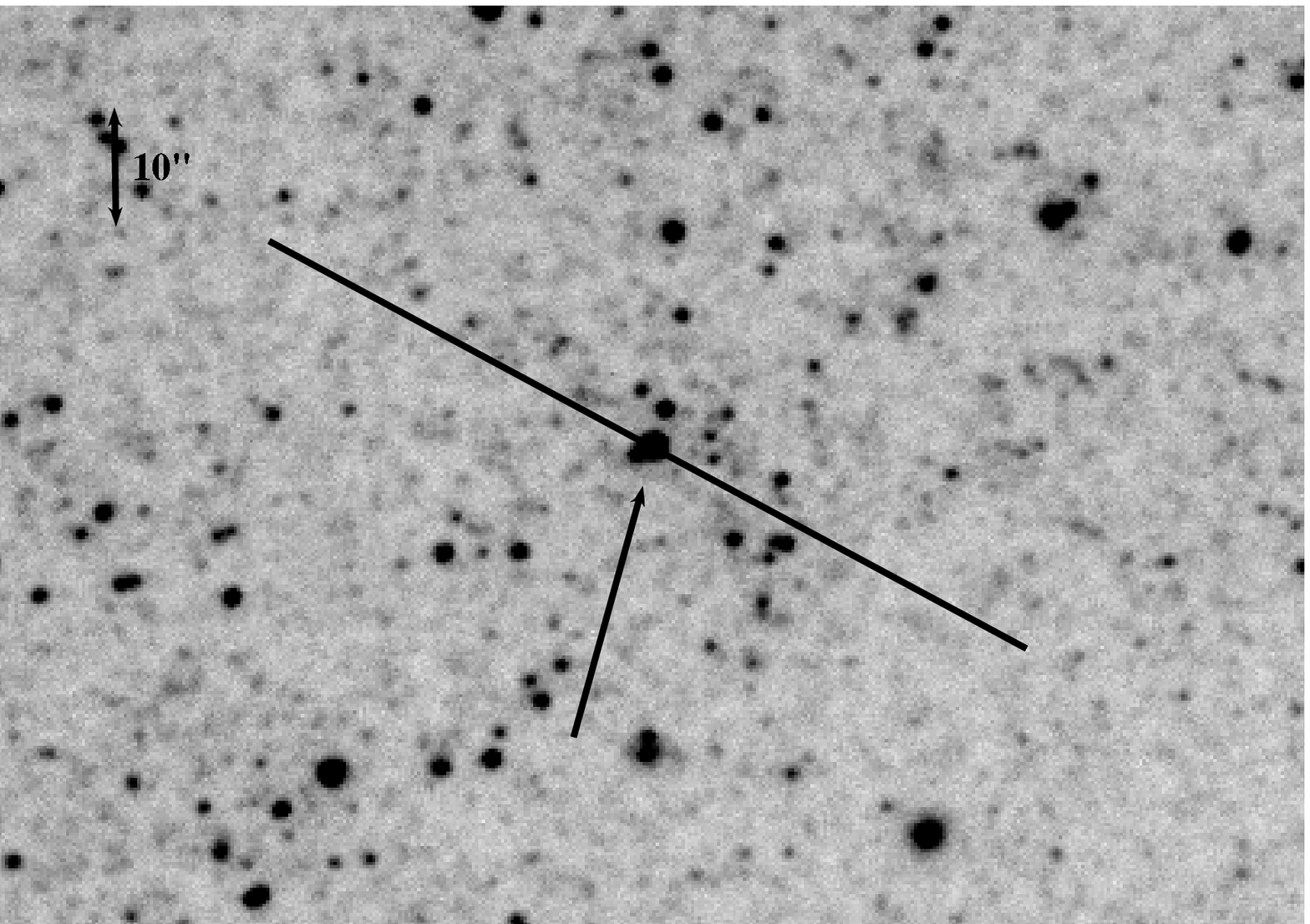}
\includegraphics[width=0.95\columnwidth]{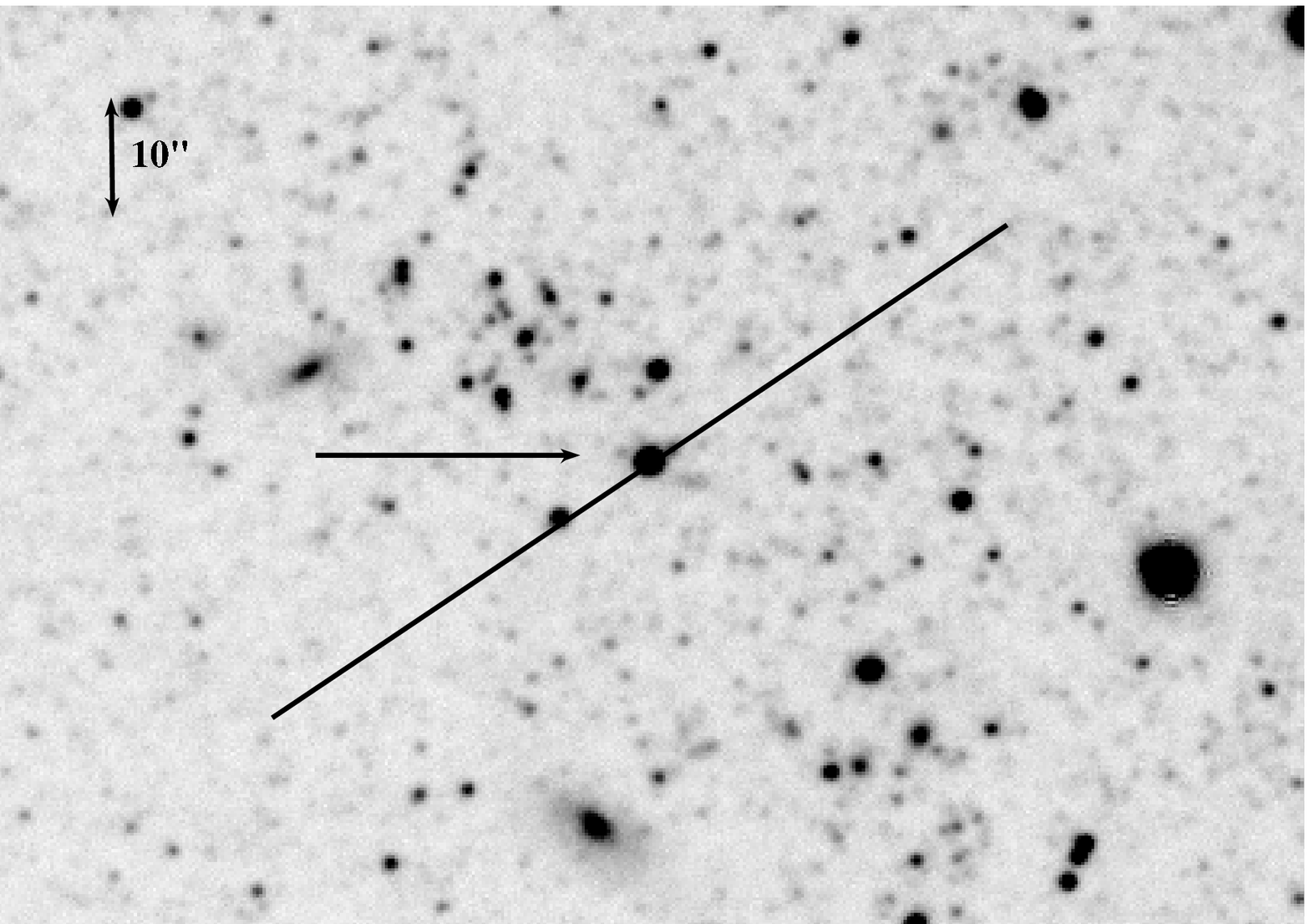}
\caption{Identification maps
in the V-band for the objects N\,45901 (top) and N\,125093
(bottom), the arrows mark their positions. The slit orientation
during the observations is shown. North is at the top, west on the
left.\label{fig:map045901}}
\end{figure}

The spectra of two new LBV candidates, N\,45901 and N\,125093 were
obtained on the 6-m BTA telescope of the Special Astrophysical
Observatory of the Russian Academy of Sciences (SAO RAS) with the
{\tt SCORPIO} focal reducer. Table \ref{table:Obs} presents the
log of observations. During the observations we used the slit
width of 1\arcsec{}, its orientation is shown in
Fig.~\ref{fig:map045901}. The quality of the images during the
observations was from 1\farcs2 to 3\farcs0.

The data processing was performed in a standard way. After the
spectra were cleaned for traces of cosmic particles, the following
reduction procedures were completed: correction for the level of
the electronic zero (bias), for flat field, wavelength calibration
based on the spectrum of a NeAr lamp, correction for spectral
sensitivity of the CCD, and reduction to fluxes in energy units.

\begin{table}
  \caption{The log of spectral observations\label{table:Obs}}
  \begin{center}
    \begin{tabular}{c|c|c|c}
\hline
Object   &  Date & Range                          & Exposure      \\
         &       & (resolution) & time \\
\hline
N\,45901& 2010.01.09 & 3700--7200~\AA (12~\AA) &  $2\times1200^s$ \\
N\,125093 & 2009.10.09 & 3700--7200~\AA (12~\AA) &  $1\times600^s$ \\
       & 2009.10.23 & 3900--5700~\AA (5~\AA) &  $2\times1200^s$\\
       & 2009.11.09 & 5700--7400~\AA (5~\AA) &  $4\times900^s$ \\
\hline
    \end{tabular}
    \end{center}

\end{table}

\subsection{Photometric Data}

We used the M\,33 images from the Spitzer Space Telescope archive
(applicant: R.H.~Gehrz), which were obtained on July 9 and August
22, 2004, and January 21 and June 10, 2005 in 4 filters (3.6, 4.5,
5.8 and 8.0~\micron) with the infrared IRAC camera
\cite{Fazio2004AboutIRACCamera}. 
The camera has separate $256\times256$ pixel CCD chips for each band with a scale of 1\farcs2 per pixel.
For the photometry we used the
images of the entire M\,33  galaxy, which are the result of
co-adding of many small-sized original images into a single image
of the galaxy with the simultaneous conversion of fluxes into
energy units.

Other Spitzer telescope archive images we used were acquired on
December 29, 2003 in the scanning mode with the MIPS instrument in
the \mbox{24~\micron{}} \mbox{filter
\cite{Rieke2004AboutMIPSCamera}.} A \mbox{$128\times128$} pixel
CCD chip with 2\farcs5 per pixel is used in this range.

%

Aperture photometry of the Spitzer images was done using the {\tt
MOPEX/APEX} codes. The aperture correction was determined from
single bright stars near the measured object. The fluxes from the
objects are listed in Table \ref{table:LBVphot}, the flux errors
do not exceed 2--3\%.

The object N\,125093 was not detectable in the MIPS camera images
in 70 and 160~\micron{} filters, but it is confidently measured in
other bands.

The measurements in the 3.6 and 4.5 \micron{} filters for the
object N\,45901 should be considered as upper limits because a
neighbouring star is in 3\farcs5. Its contribution may be
significant since the mean FWHM across the field is about 1.8
pixels, and the star is located in less than 3 pixels. The PSF
photometry may resolve the stars, but for such faint objects this
technique will only allow making a rough estimate. In the 5.8 and
8.0~\micron{} filters both the object and the nearby star get lost
in the background noise. The  object is not detectable in the
24~\micron{} images.

Photometric measurements in the J \mbox{(1.25 \micron),} H (1.65
\micron) and K (2.17 \micron) filters for N\,125093 were adopted
from the 2MASS catalog \cite{2MASS}. The images of the M\,33
galaxy in the survey were acquired in December 1997. To study the
spectral energy distribution in N\,125093, we used
non-simultaneous observations, as we have no reliable information
on the variability of this object. The 2MASS catalog has no data
on N\,45901.

In the Catalog of Variables in M\,33
\cite{Hartman2006M33Variables}  N\,45901 (this catalog lists it as
N\,241989) was suspected in variability in one of the three
MegaPrime/MegaCam filters  of the CFHT telescope (in the SDSS
i$^{\prime}$ band). In addition, it appears that N\,45901
coincides with the \Ha{} emission object candidate from the list
\cite{Calzetti1995} (its number in this list is N\,44). According
to \cite{Calzetti1995}, the magnitude of this object is V=18\fm45,
whereas in \mbox{the \cite{Massey2006UBVRIcatalog}} catalog it is
V=17\fm57. The magnitude difference  \mbox{$\Delta V \approx$
0\fm9} between two sets of observations, 1986--1987
\cite{Calzetti1995}, and 2000--2001 \cite{Massey2006UBVRIcatalog}
can hardly be explained by calibration errors. This is why it is
likely that the object N\,45901 is indeed a variable star.


The magnitude of the object N\,125093  in different catalogs
(USNO-A2.0, USNO-B1.0, NOMAD Catalog, Guide Star Catalog versions
2.2 and 2.6) varies by 0\fm65 in the V-band and by 2\fm3 in the
B-band. The differences in photometric and photographic systems
can naturally lead to a conspicuous variation in the brightness
estimates. Nevertheless, the differences are so remarkable that it
requires additional observations to verify the possible
variability.

\section{SPECTRA}

The main criterion which we used to relate these two new objects
to the class of LBV candidates are their broad and strong
emissions in \Ha~, their widths, taking into account the spectral
resolution (FWHM), are 1000 and 770\,km/s for N\,125093 and
N\,45901, respectively. Their equivalent widths are 36\AA{} and
38\AA{}. In addition, \Ha{} emissions in these stars have
significantly more extended wings. These lines are formed in the
stellar wind. The second criterion is the stellar luminosity.
Based on the spectra of these stars (see below) we can suggest the
temperature of the photosphere to be certainly higher than
10000\,K. The observed colors of stars \mbox{($B-V=$0\fm71} and
\mbox{$B-V=$0\fm85} for N\,45901 and N\,125093, respectively), as
well as the observed diffuse interstellar bands (DIBs) indicate
strong interstellar reddening, i.e. high intrinsic luminosity of
these candidates.

\begin{table}
\caption{The observed parameters of the two new LBV candidates.
The first two lines indicate the coordinates of stars. U, B, V, R,
I values were adopted from \cite{Massey2006UBVRIcatalog}, the J,
H, K data is from \cite{2MASS}, the last five lines list the
results of our photometry of the Spitzer images. The errors of
this photometry do not \mbox{exceed 2--3\,\%.}
\label{table:LBVphot}}
  \begin{center}
    \begin{tabular}{lcc}
\hline
            & N\,45901          & N\,125093\\
\hline
Ra (J2000)  & 01:33:27.40       & 01:34:15.42 \\
Dec (J2000) & +30:30:29.5       & +30:28:16.4 \\
U (mag) &18.509 $\pm$  0.004& 18.383 $\pm$ 0.004 \\
B (mag)     &18.279 $\pm$  0.004& 18.138 $\pm$ 0.004 \\ 
V (mag)     &17.572 $\pm$  0.004& 17.284 $\pm$ 0.004 \\
R (mag)     &17.193 $\pm$  0.004& 16.755 $\pm$ 0.004 \\
I (mag)     &16.807 $\pm$  0.004& 16.203 $\pm$ 0.005 \\
J (mag) & -- & 15.385 $\pm$ 0.041 \\
H (mag)     & -- & 14.819 $\pm$ 0.062 \\
K (mag)     & -- & 14.120 $\pm$ 0.048 \\
3.6\micron{} (mJy) & $3.17\times{}10^2$~$^a$ & $3.11\times{}10^3$\\
4.5\micron{} (mJy) & $1.45\times{}10^2$~$^a$& $3.69\times{}10^3$\\
5.8\micron{} (mJy) & -- &  $4.09\times{}10^3$\\
8.0\micron{} (mJy) & -- &  $6.35\times{}10^3$ \\
24.0\micron{} (mJy) & -- &  $2.00\times{}10^4$ \\
\hline
    \end{tabular}
    \end{center}
\footnotetext[1]{we consider these measurements as upper limits
due to the presence of a nearby star} \vspace{5mm}
\end{table}

\subsection{Candidate N\,125093}

The spectrum of N\,125093 is demonstrated in
Fig.\,\ref{figure:spectra} compared to the spectra of the known
LBV stars in M\,33, Var\,A, Var\,B, V\,532 and a new quite firm
LBV candidate in this galaxy, N\,93351\footnote{the spectrum of
V\,532 was kindly provided by Szeifert
\cite{Szeifert1994LBVsinM31andM33} and obtained with a resolution
of 1.2~\AA. All the remaining spectra were obtained with the
SCORPIO \cite{Valeev2009NewLBV}. The spectra of Var\,A,
Var\,B---with a spectral resolution of 12\,\AA, while the spectrum
of N\,93351---with a spectral resolution of
5\,\AA.}\cite{Valeev2009NewLBV}. The He\,I emissions in the
spectrum of this star are very weak, but noticeable, the brightest
line \mbox{He\,I $\lambda \lambda 5876$} has a low intensity. The
red region of the spectrum reveals weak Fe\,II emission lines, and
several forbidden [Fe\,II] lines. There are nebular lines,
[O\,I]\,$\lambda \lambda 6300, 6360$, and weak [N\,II]\,$\lambda
\lambda 6548, 6384$ lines in the wings of the broad \mbox{ \Ha~
emission.}

The [Ca\,II]\,$\lambda \lambda 7291, 7323$ lines are quite
interesting, they indicate a recent gas eruption and dust activity
linked with \mbox{it \cite{Valeev2009NewLBV}.} Var\,A was the
brightest star in M\,33 in 1950, and later
\cite{Humphreys1987,Humphreys2006} revealed an infrared excess and
bright [Ca\,II] lines. Typically, these lines are not observed in
the classical LBV stars, they are visible in the cold hypergiants
Var\,A in M\,33 and IRC+10420 (but the star Var\,A is a typical
LBV star based on a number of features, see
\cite{Valeev2009NewLBV}). Absorption the Fe\,II lines are visible
in the blue region of the spectrum, some Ti\,II lines (like in the
spectrum of N\,93351) and two Si\,II\,$\lambda \lambda 6347, 6371$
lines. However, the strongest metal line Fe\,II\,$\lambda 5169$
possesses emission components in the wings. In the spectrum of
N\,125093  diffuse interstellar extinction bands (DIBs) are
clearly visible, the Na\,I\,$\lambda \lambda 5890, 5896$ doublet
is as well sufficiently strong. The \Hb{} emission in the spectrum
of N\,125093 is relatively weak and does not have broad wings at
the given spectrum quality in the region.

We conducted the Gaussian analysis of the \Ha{} emission, and
found that a narrow component of this line has FWHM$=5.9$~\AA,
which is not much different from our spectral resolution \mbox{(5
\AA),} but its broad component has \mbox{FWHM$=22.5$~\AA,} which
corresponds to the velocity dispersion in the outflowing wind of
approximately $1000$\,km/s (after correcting for the spectral
resolution). The narrow component of \Hb{} line
\mbox{FWHM$=4.4$~\AA.}

Faint and extended H\,II regions were captured by the slit.
Directly around N\,125093 the H\,II emission is very weak.

To estimate the interstellar extinction, we used the \Ha{}/\Hb{}
line flux ratio  for the nebula, located near the star N\,125093.
For the gaseous nebulae this ratio amounts to approximately  2.87
in a wide range of temperatures and densities
\cite{OsterbrockFerland2006}. We used the law of interstellar
reddening from the work of \mbox{O'Donnell
\cite{ODonnell1994ReddenCurve}} at R$_V$=3.07. We obtained the
value  A$_V \lesssim$ 2.5.

A comparison of the spectrum of N\,125093 with the spectra of
relatively hot stars Var\,B and V\,532 \cite{Valeev2009NewLBV}
allows to conclude that the temperature of the photosphere of
N\,125093 is certainly lower than 20000\,K. On the other hand,
based on the presence of deep  Si\,II absorptions  we can construe
that the temperature is definitely above 10000\,K. The equivalent
width of the Si\,II $\lambda$6347 line is equal to 0.48 $\pm$ 0.02
\AA{}. In the spectra of supergiants cooler than A0, the
equivalent width of this line is equal to 0.7\AA{}
\cite{Chentsov2007SpectralAtlas,Didelon1982BstarsEW}, while in
hotter supergiants the equivalent width becomes smaller.

From the value of the equivalent width of the \mbox{Si\,II
$\lambda$6347} line, we can conclude that this is a B5--B7 star
and its temperature is $>$ 10000~K. The spectrum of this star is
very similar to the spectrum of  N\,93351, the temperature of
which we earlier estimated \cite{Valeev2009NewLBV} as
13000--16000\,K. Hence we can make a rough estimate of the
temperature of N\,125093 from its spectral lines, \mbox{T $\sim
12000 - 16000$\,K.} We will later use these preliminary estimates
of the temperature along with that of interstellar extinction in
the spectral energy distribution fit of N\,125093, where we will
specify both the temperature and  reddening.

\subsection{Candidate N\,45901}

\begin{figure*}[tbp]
 \includegraphics[width=7cm,angle=270]{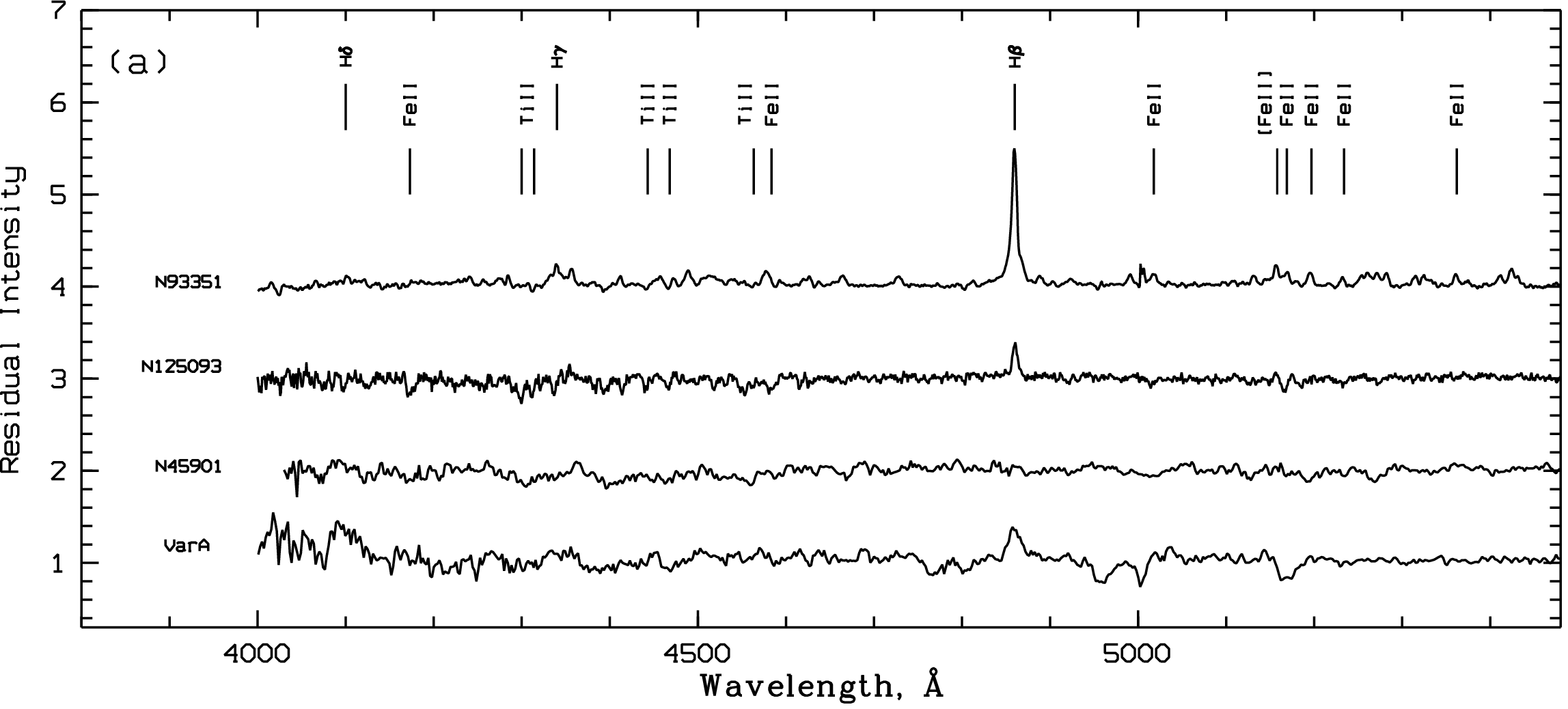}
 \includegraphics[width=7cm,angle=270]{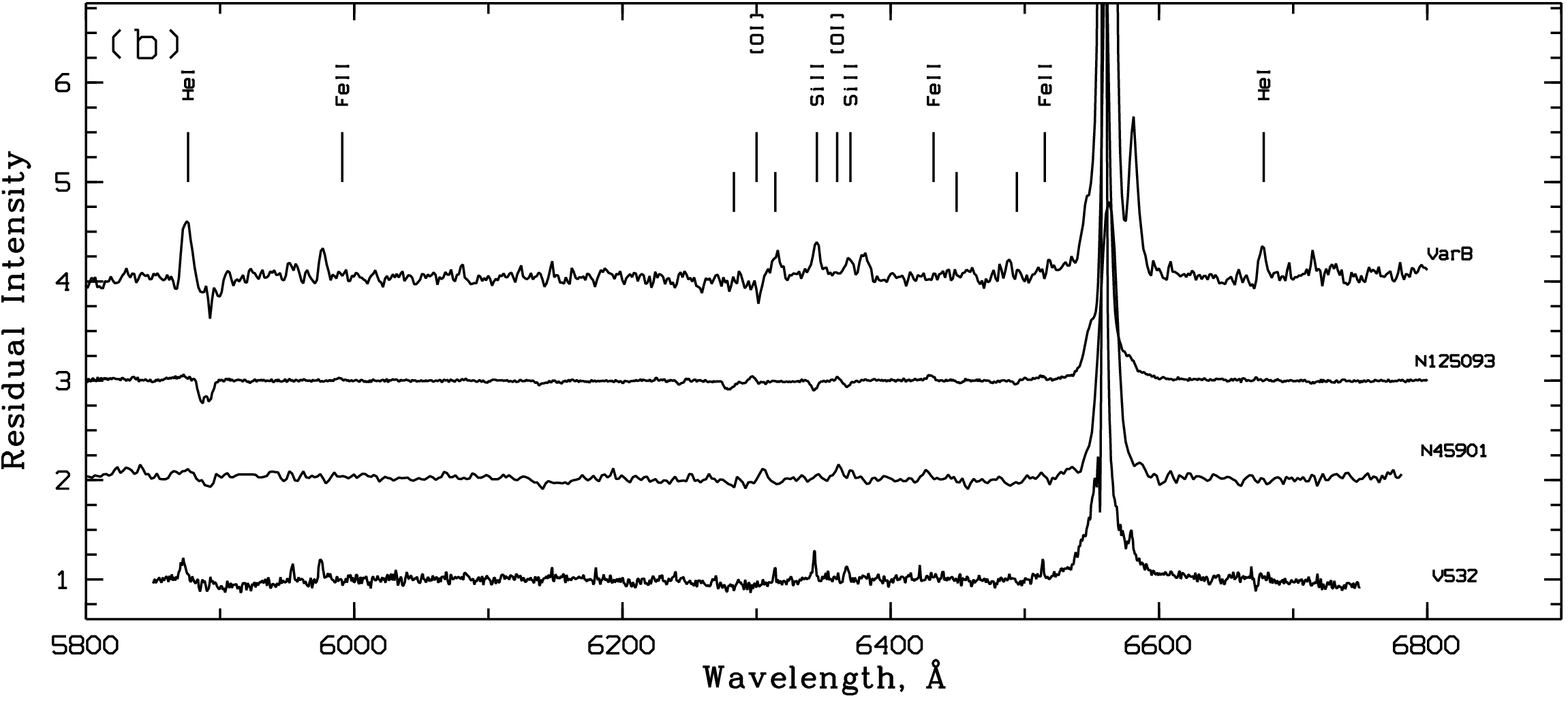}
 \includegraphics[width=7cm,angle=270]{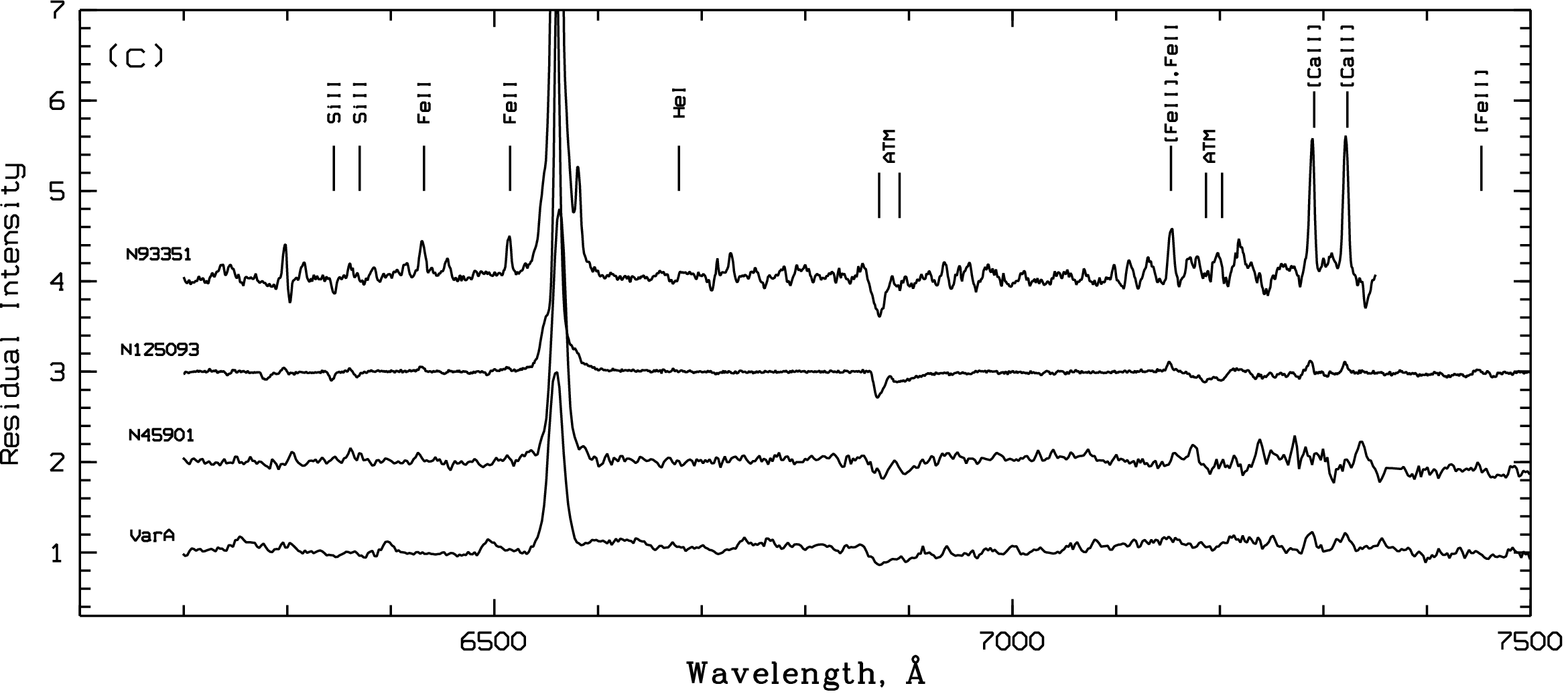}
\caption{
The spectra of LBV candidates N\,45901 and N\,125093 along with
the spectra of the known LBV in the M\,33 galaxy, Var\,A (an LBV or a red hypergiant) and N\,93351 (a
firm LBV candidate) (a, c), Var\,B and
V\,532 (b).
The spectra of both new LBV candidates  reveal broad \Ha{}
emission lines. The main spectral lines detected in the spectra of
candidates are marked.
\label{figure:spectra}}
\end{figure*}

The spectrum of the star N\,45901 is shown in
Fig.\,\ref{figure:spectra}.  Unfortunately, due to weather
conditions we were unable to obtain the spectrum of this star with
a resolution of 5\,\AA{}, and even the existing spectrum with a
resolution of  12\,\AA{} has a low signal-to-noise ratio.
Nevertheless, there is a bright and broad \Ha{} emission with
extended wings in the spectrum of N\,45901. Gaussian analysis
revealed a two-component line profile: a bright narrow line and
broad wings. The widths of the narrow and broad components,
corrected for spectral resolution, are equal to 140~km/s and
770~km/s, respectively. There is a [N\,II]\,$\lambda 6584$ line in
the broad \Ha{} wing. The second line of the [N\,II] doublet,
which should be 3 times weaker than the first, is probably lost in
the bright blue \Ha{} wing. We also found forbidden lines
[O\,I]\,$\lambda \lambda 6300, 6360$, which are formed in the
unresolved region around the star sized smaller than 4--5 pc (at
the distance we adopted, the scale in M\,33 is equal to
4.6\,pc/$''$). Despite the poor quality of the spectrum,
especially in the blue region, the absorption lines Fe\,II and
Ti\,II that are among the most intensive lines of these ions, are
visible in this range. In the red range, we may even suspect the
[Fe\,II] emissions. The region shorter than the \Ha{} line reveals
signatures of two strongest Fe\,II emission lines.

Based on the presence of a broad and bright \Ha{} emission, as
well as the likely occurrence of Fe\,II and Ti\,II lines we can
make a rough estimate that the temperature of the photosphere of
N\,45901 is above 10000--12000\,K. We did not detect any He\,I
lines, hence its temperature is below 15000\,K. Neither did we
find any H\,II regions around the star N\,45901, which might be
due to the low quality of the spectrum. This is why we shall
continue estimating the temperature of the star and the value of
interstellar extinction from its spectral energy distribution
without preliminary evaluations of reddening.

\begin{table}
\caption{SED modelling results. Two limiting versions of
permissible values of interstellar reddening, and the
corresponding bolometric luminosities and temperatures of the hot
components (stars) in Kelvins are listed. The last two columns
indicate the temperatures of warm and cold dust components, and
their percentage contribution to the bolometric luminosity in the
parentheses. \label{table:LBVparam}}
  \begin{center}
    \begin{tabular}{lllllll}
\hline
Object   & $A_V$  &$\log$L/L$_{\sun}$ &T$_{\mbox{*}}$& T$_{\mbox{warm}}$      & T$_{\mbox{cold}}$& \\
\hline
N\,45901 &    2.2 &   6.02 &         13200   &   1000(0.1) & & \\
         &    2.4 &   6.21 &         15200   &   1000(0.1) & & \\
N\,125093&    2.6 &   6.30 &  12700 &   1400(3.9) &    480(3.2) & \\
         &    2.9 &   6.60 &  16200 &   1400(2.0) &    470(1.6) & \\
\hline
    \end{tabular}
    \end{center}
\end{table}

\section{SPECTRAL ENERGY DISTRIBUTIONS  AND STELLAR PARAMETERS}

Before refining the temperatures and luminosities of these stars
in their energy distributions, we estimate their interstellar
reddening. It follows from the spectra that the temperatures of
the photospheres of these stars are above 10000\,K. For the
supergiants this temperature corresponds to \mbox{$(B - V)_0 =
0\fm0$} at the bolometric correction \mbox{$BC = -0\fm3$
\cite{Flower1996BolometricCorrection}.} The observed colors of the
stars are equal to \mbox{$(B - V) = 0\fm71$} for N\,45901 and
\mbox{$(B - V) = 0\fm85$} for N\,125093. Therefore, their values
of interstellar reddening are not below \mbox{$A_V = 2.2$} and
2.6, respectively. At the distance modulus to M\,33 we adopted as
equal to \mbox{$(m - M)_0 = 24\fm9$
\cite{Bananos2006M33Distance}}, we find the lower limit of
N\,45901 luminosity as \mbox{$M_V = -9\fm5$,} \mbox{$M_{bol} =
-9\fm8$} and \mbox{$M_V = -9\fm9$,} \mbox{$M_{bol} = -10\fm2$} for
N\,125093.

To construct the spectral energy distributions (SED), we used the
data of optical and infrared photometry from
Table\,\ref{table:LBVphot}. We applied a blackbody approximation
for individual components: a hot star and one or two components of
dust emission. The upper limit of the temperature of the warm dust
component was set at 1500~K, since above this temperature the dust
begins to evaporate.

Without the a priori information on temperature and reddening it
is not possible to determine these two parameters independently.
For the object N\,125093, we have both a preliminary estimate of
the stellar temperature, and the reddening
(\mbox{13000--16000\,K,} \mbox{A$_V \lesssim 2.5$}). For N\,45901,
we only have an evaluation of the stellar photosphere temperature
\mbox{(12000--15000\,K)}. Indeed, the spectra of both our stars
are very similar (Fig.\ref{figure:spectra}), hence, their
temperatures should not differ greatly either.

\begin{figure}[h]
\includegraphics[bb=28 2 416 208, clip, width=\columnwidth]{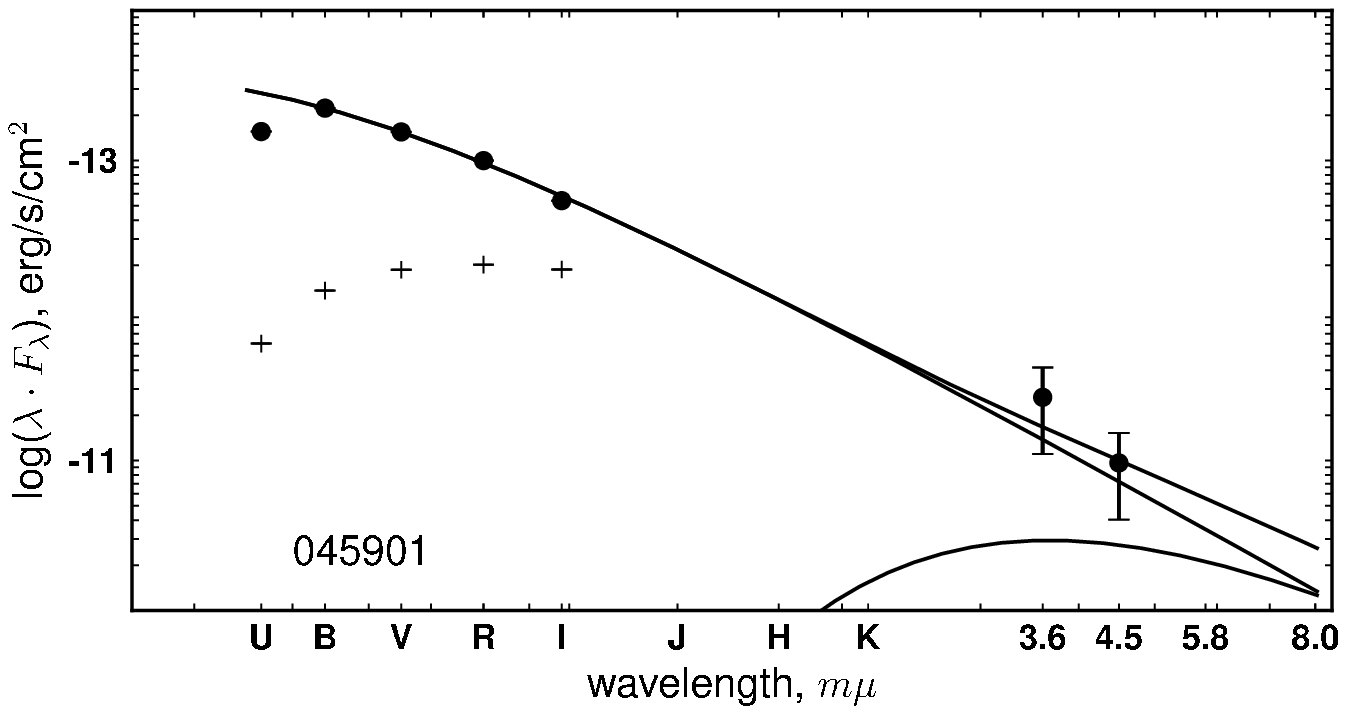}
\includegraphics[bb=28 2 416 208, clip, width=\columnwidth]{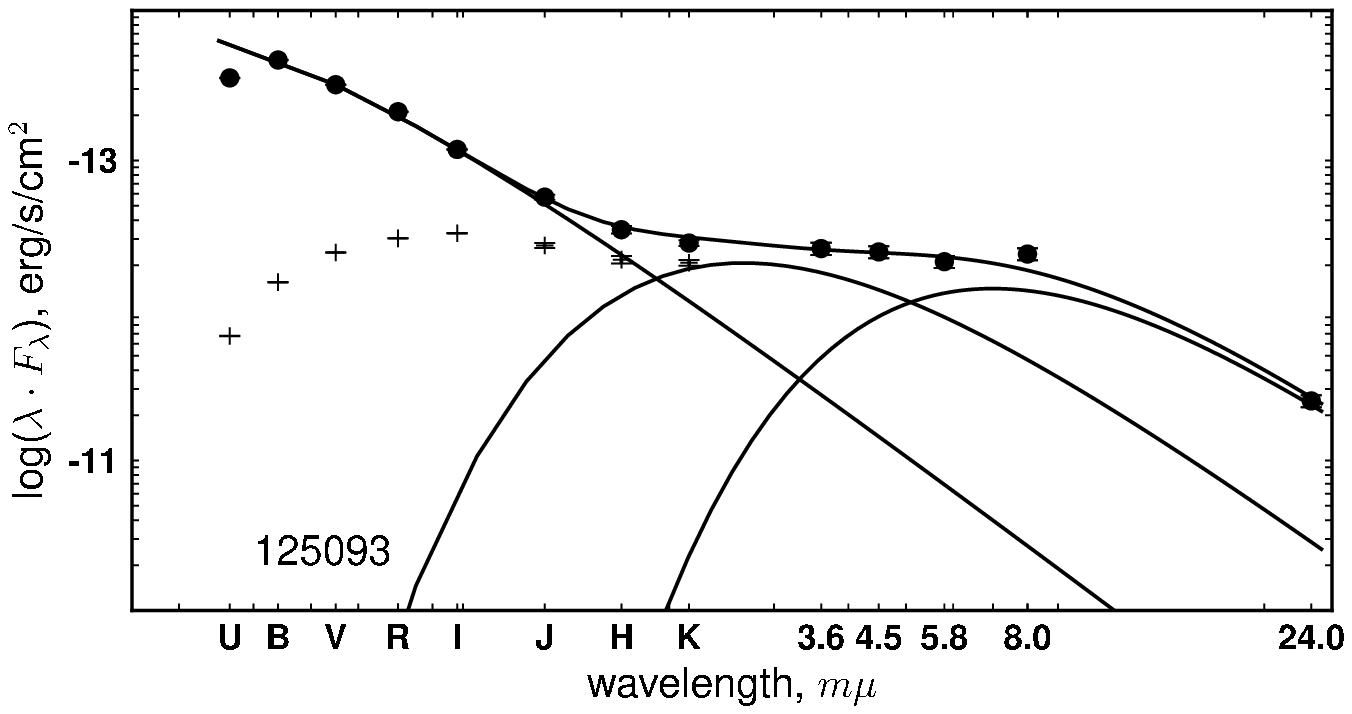}
\caption{ SEDs of two new LBV candidates in
M\,33. The results of photometry not corrected for reddening are
marked with crosses. The corrected measurements (for N\,45901
$A_{V}=2.3$ and for N\,125093 $A_{V}=2.8$)  are indicated by
circles. The errors represent the accuracy of photometry. If the
size of the circle is larger than the measurement error, the error
bars are not visible. The solid lines represent individual
blackbody components of radiation, as well as the sum of all
components. \label{figure:SEDs}}
\end{figure}

The SEDs are demonstrated in Fig.\,\ref{figure:SEDs}. The
radiation intensity drop in the U-band is associated with the
presence of the Balmer jump in these stars. The magnitude of
fading in the U-band is about the same for both stars. This
independently confirms our conclusion that the temperatures of
both stars are almost identical. We have not used the measurements
obtained in the U-band at the fits. We assume that the photometric
measurements in the R, I (and, possibly, J-bands) may be distorted
by an additional contribution of the free-free wind radiation
\cite{Valeev2009NewLBV}. However, the result of approximations
shows that the contribution of this emission in the two stars is
negligible.

The luminosity of each SED component  was calculated as the area
under the corresponding blackbody spectrum, and the bolometric
luminosity of the object was deduced as the sum of luminosities of
the components. The best solutions for these stars are listed in
Table \ref{table:LBVparam}. The photometric data (not corrected
for reddening) are marked with crosses in Fig.~\ref{figure:SEDs}.
The measurements corrected for reddening (\mbox{A$_{V}=2.8$} for
N\,125093, and \mbox{A$_{V}=2.3$} for N\,45901) are indicated by
circles. The errors represent the accuracy of photometry. If the
size of the circle is larger than the measurement error, the error
bars are not visible. The solid lines represent individual
blackbody components (three for N\,125093, and two for N\,45901),
as well as the sum of all components.

For N\,125093 we find the optimal value of the interstellar
extinction as \mbox{A$_V = 2.75 \pm 0.15$.} It is close to the
assessment we obtained from the spectra of the surrounding H\,II
emission (A$_V \lesssim 2.5$). The stellar parameters obtained for
the limiting values of A$_V$=2.6 and A$_V$=2.9 are listed in
Table\,\ref{table:LBVparam}. The temperature of the stellar
photosphere is \mbox{T$_\star \sim 13000 - 16000$\,K,} which is
consistent with our estimate from the spectrum. The corresponding
bolometric luminosity of  N\,125093 amounts to
$\log$(L/L$_{\sun}$) = 6.3 -- 6.6. We find two thermal components
in the spectrum, a warm component with \mbox{T$_{\mbox{warm}} \sim
1000$\,K,} and a cold component with \mbox{T$_{\mbox{cold}} \sim
480$\,K.} The infrared excess in N\,125093 is 5--6\,\% of the
bolometric luminosity.

In the case of N\,45901 we deduced the optimal estimate of
interstellar extinction  at A$_V = 2.3 \pm 0.1$. Accordingly, the
temperature of the photosphere of N\,45901 is estimated as
\mbox{T$_{\star} \sim$ 13000--15000\,K,}
\linebreak and its bolometric luminosity   amounts to \linebreak
\mbox{$\log$(L/L$_{\sun}$) = 6.0 -- 6.2.} Infrared excess in
N\,45901 corresponds to the warm dust radiation with the
temperature of \mbox{T$_{\mbox{warm}} \sim 1000$\,K} and amounts
to 0.1\,\% of the bolometric luminosity.

\begin{figure*}[tbp]
\includegraphics[width=0.8\textwidth]{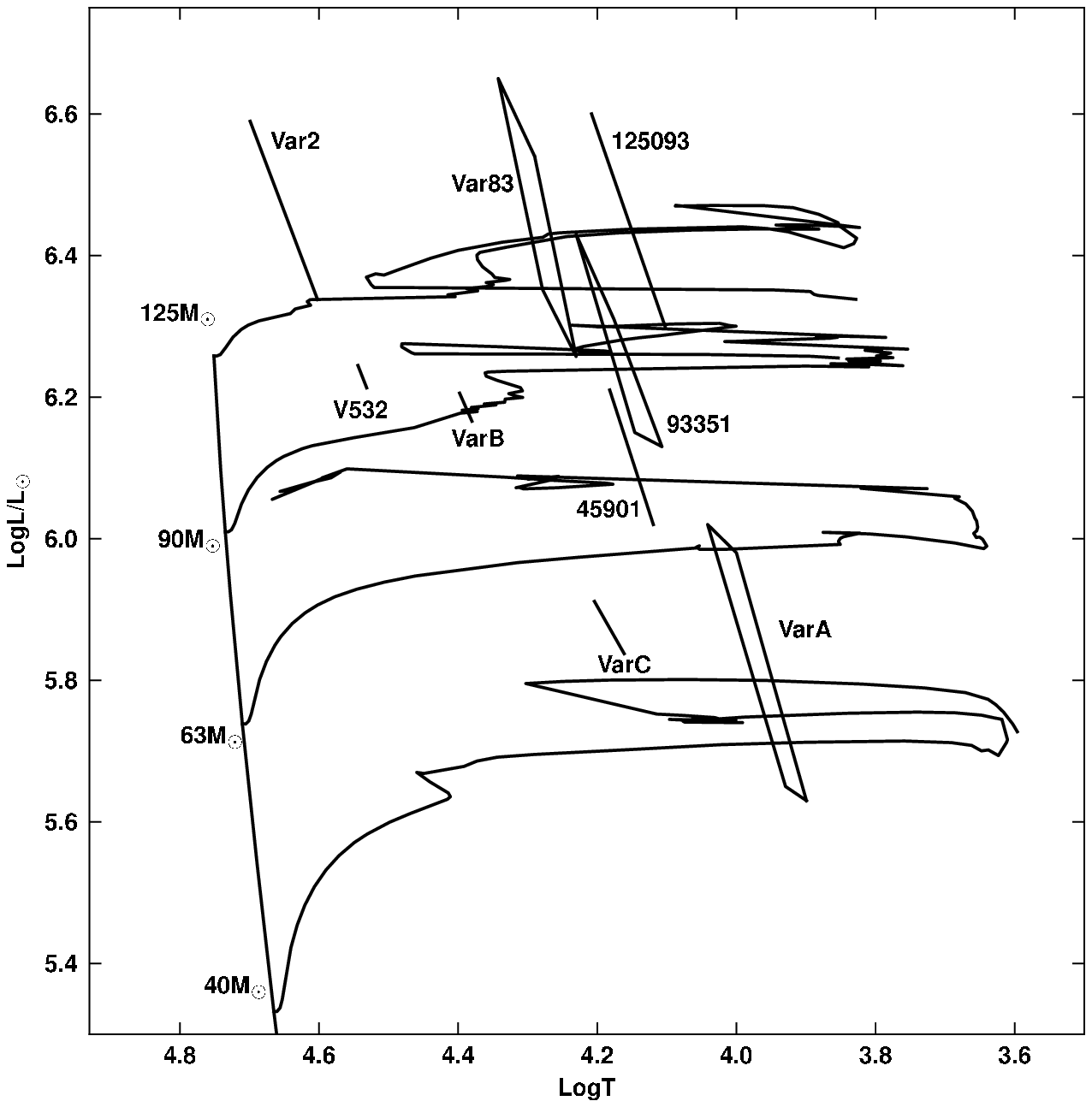}
\caption{The temperature--luminosity diagram
for all the LBV stars known to date (Var\,B, Var\,C, Var\,2,
Var\,83, V\,532), as well as the stars Var\,A and N\,93351 from
the M\,33 galaxy (data adopted from \cite{Valeev2009NewLBV}), with
two new candidates N\,45901 and N\,125093 imposed. The evolutional
tracks,  computed for metallicity Z=0.007 are adopted from
\cite{Claret2006Evolution0.007}.\label{figure:HR}}
\end{figure*}

Figure\,\ref{figure:HR} demonstrates the  temperature--luminosity
diagram for all the LBV stars known to date (Var\,B, Var\,C,
Var\,2, Var\,83, V\,532), as well as the stars Var\,A and N\,93351
in the M\,33 galaxy, the parameters of which were determined in
our paper \cite{Valeev2009NewLBV}. We show there our two new
candidates N\,45901 and N\,125093, and the evolutional tracks
\cite{Claret2006Evolution0.007}, computed for the stars with
metallicity 0.007, corresponding to M\,33. The temperatures and
luminosities of these stars were deduced by a unified method of
approximation of the SEDs engaging additional data on the
temperatures of stars and interstellar extinction values, obtained
from the spectra. For the majority of stars the boundary values of
temperature are shown, which have met the approximation of
spectral energy distributions. In the case of stars Var\,83,
Var\,A, and N\,93351, the parameters of the two boundary values of
interstellar extinction are indicated, hence the range of valid
parameters (L,T) forms a closed region.

From this diagram, we find the mass estimates for the stars on the
ZAMS. The masses of N\,45901 and N\,125093 are estimated as
approximately \mbox{$60-80 M_{\sun}$,} and \mbox{$90-120
M_{\sun}$,} respectively.

\section{CONCLUSION}

Luminous blue variables are characterized by a strong and diverse
brightness  \mbox{variability \cite{vanGenderen2001}.} A large
amplitude variability (over 1$^m$) is rare in the classical LBVs,
such events may take place on the time scales of dozens of years.
For example, an LBV prototype P\,Cyg, having demonstrated a giant
eruption, possesses a relatively stable brightness over the past
three centuries. In fact, any new object candidate for the LBV
will remain only an ``LBV candidate'' until its brightness
variability reaches a large amplitude (presumably not less than
0.3--0.5 magnitude, as the amplitudes of 0.1--0.2 magnitude is
common for massive supergiants). Tens and hundreds of years may
pass while we wait for a strong manifestation of brightness
variability. Regardless of the spectrum, temperature, or
luminosity, irrespective of which ``typical LBV'' features are
found in one or another LBV candidate, such an object will remain
only a candidate.

We hope that this state of physics of LBV objects will not persist
long. There will be found more reliable criteria and
classifications of the LBV class and this stage of massive star
evolution in general. The chemical composition of the atmosphere,
i.e. the relative hydrogen and helium abundance is likely to be
such a reliable criterion. The measurements of abundances require
spectral analysis based on the models of expanding atmospheres.
Indeed, the LBV star well-known in our Galaxy, AG\,Car has the
abundance He/H$=0.43$, which was deduced via detailed modelling of
its spectrum \cite{Groh2009AGCarNature}. A relation between the
LBV stars and late WR stars of the nitrogen sequence with hydrogen
in the atmospheres (WNLh, see the Introduction), which reveal a
low abundance of hydrogen in the wind is well-known. It is obvious
that the understanding of physics and evolutionary status of the
LBV objects requires detecting large numbers of objects bearing
properties similar to LBV.

This paper presents a preliminary study of two new LBV candidates
in the M\,33 galaxy, which were discovered by us in the spectral
observations of a list of emission object candidates with a
presumably notable interstellar reddening
\cite{ValeevCatalogLBV2010}. The spectra of N\,45901 and N\,125093
are similar to the spectra of the stars at the LBV stage. They
have strong and broad \Ha, emissions, forbidden [O\,I] and [N\,II]
lines. The spectrum of the second star reveals [Ca\,II], [Fe\,II]
and Fe\,II emission lines, Ti\,II and Fe\,II absorptions. The
spectrum of the first star has an insufficient quality for a
reliable identification, however the [O\,I], Fe\,II and Ti\,II
lines are present in its spectrum. Additional spectra with a
better resolution and signal-to-noise ratio are required to
accurately examine the spectrum. Comparisons of brightness
estimates from different catalogs indicate a probable variability
of the object N\,45901.

We found an infrared excess in both stars. The  temperatures of
warm and cold dust components in the star N\,125093 (it possesses
the [Ca\,II] emissions), amount to 1400\,K and 470\,K,
respectively.

We estimate that the star N\,45901 has a bolometric luminosity of
$\log$(L/L$_{\sun}$) = 6.0 -- 6.2 and its probable mass on the
initial main sequence is \mbox{M $\sim 60-80$ M$_{\sun}$.} The
luminosity of N\,125093 amounts to  \mbox{$\log$(L/L$_{\sun}$) =
6.3 -- 6.6} and its initial mass is  \mbox{M $\sim 90-120
$M$_{\sun}$.} All the above properties of  N\,45901 and N\,12509
allow us to classify them as LBV candidates.

\begin{acknowledgments}
The authors thank T.A.~Fatkhullin for help in observations, and
E.L.~Chentsov for the useful discussions of spectral
classifications. This paper was supported by the grants of the
Russian Foundation for Basic Research (\No{} 09-02-00163
and 10-02-00463), and the grant ``Leading Scientific Schools of
Russia'' \No{} 5473.2010.2.
\end{acknowledgments}


\begin{thebibliography}{32}
\expandafter\ifx\csname
natexlab\endcsname\relax\def\natexlab#1{#1}\fi
\expandafter\ifx\csname bibnamefont\endcsname\relax
  \def\bibnamefont#1{#1}\fi
\expandafter\ifx\csname bibfnamefont\endcsname\relax
  \def\bibfnamefont#1{#1}\fi
\expandafter\ifx\csname citenamefont\endcsname\relax
  \def\citenamefont#1{#1}\fi
\expandafter\ifx\csname url\endcsname\relax
  \def\url#1{\texttt{#1}}\fi
\expandafter\ifx\csname
urlprefix\endcsname\relax\def\urlprefix{URL }\fi
\providecommand{\bibinfo}[2]{#2}
\providecommand{\eprint}[2][]{\url{#2}}

\bibitem[{\citenamefont{{Valeev} et~al.}(2010)\citenamefont{{Valeev},
  {Sholukhova}, and {Fabrika}}}]{ValeevCatalogLBV2010}
\bibinfo{author}{\bibfnamefont{A.~F.} \bibnamefont{{Valeev}}},
  \bibinfo{author}{\bibfnamefont{O.}~\bibnamefont{{Sholukhova}}},
  \bibnamefont{and}
  \bibinfo{author}{\bibfnamefont{S.}~\bibnamefont{{Fabrika}}},
  \bibinfo{journal}{\ab} \textbf{\bibinfo{volume}{65}}, \bibinfo{pages}{140}
  (\bibinfo{year}{2010}).


\bibitem[{\citenamefont{{Massey} et~al.}(2006)\citenamefont{{Massey}, {Olsen}, {Hodge},{Strong}, {Jacoby}, {Schlingman}, and {Smith}}}]{Massey2006UBVRIcatalog}
\bibinfo{author}{\bibfnamefont{P.}~\bibnamefont{{Massey}}},
  \bibinfo{author}{\bibfnamefont{K.~A.~G.} \bibnamefont{{Olsen}}},
  \bibinfo{author}{\bibfnamefont{P.~W.} \bibnamefont{{Hodge}}},
et~al.,
 \bibinfo{journal}{\aj}
  \textbf{\bibinfo{volume}{131}}, \bibinfo{pages}{2478} (\bibinfo{year}{2006}).

\bibitem[{\citenamefont{{Humphreys} and
  {Davidson}}(1994)}]{HumphreysDavidson1994}
\bibinfo{author}{\bibfnamefont{R.~M.} \bibnamefont{{Humphreys}}}
  \bibnamefont{and}
  \bibinfo{author}{\bibfnamefont{K.}~\bibnamefont{{Davidson}}},
  \bibinfo{journal}{\pasp} \textbf{\bibinfo{volume}{106}},
  \bibinfo{pages}{1025} (\bibinfo{year}{1994}).

\bibitem[{\citenamefont{{Meynet} et~al.}(2007)\citenamefont{{Meynet},
  {Eggenberger}, and {Maeder}}}]{Meynet2007}
\bibinfo{author}{\bibfnamefont{G.}~\bibnamefont{{Meynet}}},
  \bibinfo{author}{\bibfnamefont{P.}~\bibnamefont{{Eggenberger}}},
  \bibnamefont{and} \bibinfo{author}{\bibfnamefont{A.}~\bibnamefont{{Maeder}}},
  \bibinfo{journal}{IAU Symposium}  \textbf{\bibinfo{volume}{241}},
 \bibinfo{pages}{13} (\bibinfo{year}{2006}).


\bibitem[{\citenamefont{{Smith} and {Conti}}(2008)}]{SmithConti2008}
\bibinfo{author}{\bibfnamefont{N.}~\bibnamefont{{Smith}}} \bibnamefont{and}
  \bibinfo{author}{\bibfnamefont{P.~S.} \bibnamefont{{Conti}}},
  \bibinfo{journal}{\apj} \textbf{\bibinfo{volume}{679}}, \bibinfo{pages}{1467}
  (\bibinfo{year}{2008}).

\bibitem[{\citenamefont{{Fabrika} et~al.}(2005)\citenamefont{{Fabrika},
  {Sholukhova}, {Becker}, {Afanasiev}, {Roth}, and {Sanchez}}}]{Fabrika2005}
\bibinfo{author}{\bibfnamefont{S.}~\bibnamefont{{Fabrika}}},
  \bibinfo{author}{\bibfnamefont{O.}~\bibnamefont{{Sholukhova}}},
  \bibinfo{author}{\bibfnamefont{T.}~\bibnamefont{{Becker}}},
et~al.,
  \bibinfo{journal}{\aap} \textbf{\bibinfo{volume}{437}}, \bibinfo{pages}{217}
  (\bibinfo{year}{2005}).

\bibitem[{\citenamefont{{Koenigsberger}
  et~al.}(2010)\citenamefont{{Koenigsberger}, {Georgiev}, {Hillier}, {Morrell},
  {Barb{\'a}}, and {Gamen}}}]{Koenigsberger2010}
\bibinfo{author}{\bibfnamefont{G.}~\bibnamefont{{Koenigsberger}}},
  \bibinfo{author}{\bibfnamefont{L.}~\bibnamefont{{Georgiev}}},
  \bibinfo{author}{\bibfnamefont{D.~J.} \bibnamefont{{Hillier}}},
et~al.,
  \bibinfo{journal}{\aj} \textbf{\bibinfo{volume}{139}}, \bibinfo{pages}{2600}
  (\bibinfo{year}{2010}).

\bibitem[{\citenamefont{{van Genderen}}(2001)}]{vanGenderen2001}
\bibinfo{author}{\bibfnamefont{A.~M.} \bibnamefont{{van Genderen}}},
  \bibinfo{journal}{\aap} \textbf{\bibinfo{volume}{366}}, \bibinfo{pages}{508}
  (\bibinfo{year}{2001}).

\bibitem[{\citenamefont{{Gvaramadze}
  et~al.}(2010{\natexlab{a}})\citenamefont{{Gvaramadze}, {Kniazev}, and
  {Fabrika}}}]{Gvaramadze2010SpitzerSearch}
\bibinfo{author}{\bibfnamefont{V.~V.} \bibnamefont{{Gvaramadze}}},
  \bibinfo{author}{\bibfnamefont{A.~Y.} \bibnamefont{{Kniazev}}},
  \bibnamefont{and}
  \bibinfo{author}{\bibfnamefont{S.}~\bibnamefont{{Fabrika}}},
  \bibinfo{journal}{\mnras} \textbf{\bibinfo{volume}{405}},
  \bibinfo{pages}{1047} (\bibinfo{year}{2010}{\natexlab{a}}).

\bibitem[{\citenamefont{{Gvaramadze}
  et~al.}(2010{\natexlab{b}})\citenamefont{{Gvaramadze}, {Kniazev}, {Fabrika},
  {Sholukhova}, {Berdnikov}, {Cherepashchuk}, and
  {Zharova}}}]{Gvaramadze2010LBVcand}
\bibinfo{author}{\bibfnamefont{V.~V.} \bibnamefont{{Gvaramadze}}},
  \bibinfo{author}{\bibfnamefont{A.~Y.} \bibnamefont{{Kniazev}}},
  \bibinfo{author}{\bibfnamefont{S.}~\bibnamefont{{Fabrika}}},
et~al.,
\bibinfo{journal}{\mnras}
  \textbf{\bibinfo{volume}{405}}, \bibinfo{pages}{520}
  (\bibinfo{year}{2010}{\natexlab{b}}).

\bibitem[{\citenamefont{{Zaritsky} et~al.}(1989)\citenamefont{{Zaritsky},
  {Elston}, and {Hill}}}]{Zaritsky1989M33Inclination}
\bibinfo{author}{\bibfnamefont{D.}~\bibnamefont{{Zaritsky}}},
  \bibinfo{author}{\bibfnamefont{R.}~\bibnamefont{{Elston}}}, \bibnamefont{and}
  \bibinfo{author}{\bibfnamefont{J.~M.} \bibnamefont{{Hill}}},
  \bibinfo{journal}{\aj} \textbf{\bibinfo{volume}{97}}, \bibinfo{pages}{97}
  (\bibinfo{year}{1989}).

\bibitem[{\citenamefont{{Bonanos} et~al.}(2006)\citenamefont{{Bonanos},
  {Stanek}, {Kudritzki}, {Macri}, {Sasselov}, {Kaluzny}, {Stetson}, {Bersier},
  {Bresolin}, {Matheson} et~al.}}]{Bananos2006M33Distance}
\bibinfo{author}{\bibfnamefont{A.~Z.} \bibnamefont{{Bonanos}}},
  \bibinfo{author}{\bibfnamefont{K.~Z.} \bibnamefont{{Stanek}}},
  \bibinfo{author}{\bibfnamefont{R.~P.} \bibnamefont{{Kudritzki}}},
et~al.,
\bibinfo{journal}{\apj} \textbf{\bibinfo{volume}{652}},
  \bibinfo{pages}{313} (\bibinfo{year}{2006}).

\bibitem[{\citenamefont{{Clark} et~al.}(2005)\citenamefont{{Clark}, {Larionov},
  and {Arkharov}}}]{Clark2005}
\bibinfo{author}{\bibfnamefont{J.~S.} \bibnamefont{{Clark}}},
  \bibinfo{author}{\bibfnamefont{V.~M.} \bibnamefont{{Larionov}}},
  \bibnamefont{and}
  \bibinfo{author}{\bibfnamefont{A.}~\bibnamefont{{Arkharov}}},
  \bibinfo{journal}{\aap} \textbf{\bibinfo{volume}{435}}, \bibinfo{pages}{239}
  (\bibinfo{year}{2005}).

\bibitem[{\citenamefont{{Massey} et~al.}(2007)\citenamefont{{Massey},
  {McNeill}, {Olsen}, {Hodge}, {Blaha}, {Jacoby}, {Smith}, and
  {Strong}}}]{Massey2007LBVcatalog}
\bibinfo{author}{\bibfnamefont{P.}~\bibnamefont{{Massey}}},
  \bibinfo{author}{\bibfnamefont{R.~T.} \bibnamefont{{McNeill}}},
  \bibinfo{author}{\bibfnamefont{K.~A.~G.} \bibnamefont{{Olsen}}},
et~al.,
\bibinfo{journal}{\aj}
  \textbf{\bibinfo{volume}{134}}, \bibinfo{pages}{2474} (\bibinfo{year}{2007}).

\bibitem[{\citenamefont{{Fabrika} and {Sholukhova}}(1999)}]{Fabrika1999}
\bibinfo{author}{\bibfnamefont{S.}~\bibnamefont{{Fabrika}}} \bibnamefont{and}
  \bibinfo{author}{\bibfnamefont{O.}~\bibnamefont{{Sholukhova}}},
  \bibinfo{journal}{\aaps} \textbf{\bibinfo{volume}{140}}, \bibinfo{pages}{309}
  (\bibinfo{year}{1999}).

\bibitem[{\citenamefont{{Allen}}(1977)}]{Straizhis}
\bibinfo{author}{\bibfnamefont{K.~U.} \bibnamefont{{Allen}}},
  \emph{\bibinfo{title}{Astrofizicheskie velichini}} (\bibinfo{publisher}{Mir, Moscow}, \bibinfo{year}{1977}) [in Russian].

\bibitem[{\citenamefont{{Valeev} et~al.}(2009)\citenamefont{{Valeev},
  {Sholukhova}, and {Fabrika}}}]{Valeev2009NewLBV}
\bibinfo{author}{\bibfnamefont{A.~F.} \bibnamefont{{Valeev}}},
  \bibinfo{author}{\bibfnamefont{O.}~\bibnamefont{{Sholukhova}}},
  \bibnamefont{and}
  \bibinfo{author}{\bibfnamefont{S.}~\bibnamefont{{Fabrika}}},
  \bibinfo{journal}{\mnras} \textbf{\bibinfo{volume}{396}},
  \bibinfo{pages}{L21} (\bibinfo{year}{2009}).

\bibitem[{\citenamefont{{Fazio} et~al.}(2004)\citenamefont{{Fazio}, {Hora},
  {Allen}, {Ashby}, {Barmby}, {Deutsch}, {Huang}, {Kleiner}, {Marengo},
  {Megeath} et~al.}}]{Fazio2004AboutIRACCamera}
\bibinfo{author}{\bibfnamefont{G.~G.} \bibnamefont{{Fazio}}},
  \bibinfo{author}{\bibfnamefont{J.~L.} \bibnamefont{{Hora}}},
  \bibinfo{author}{\bibfnamefont{L.~E.} \bibnamefont{{Allen}}},
et~al.,
\bibinfo{journal}{\apjs}
  \textbf{\bibinfo{volume}{154}}, \bibinfo{pages}{10} (\bibinfo{year}{2004}).

\bibitem[{\citenamefont{{Rieke} et~al.}(2004)\citenamefont{{Rieke}, {Young},
  {Engelbracht}, {Kelly}, {Low}, {Haller}, {Beeman}, {Gordon}, {Stansberry},
  {Misselt} et~al.}}]{Rieke2004AboutMIPSCamera}
\bibinfo{author}{\bibfnamefont{G.~H.} \bibnamefont{{Rieke}}},
  \bibinfo{author}{\bibfnamefont{E.~T.} \bibnamefont{{Young}}},
  \bibinfo{author}{\bibfnamefont{C.~W.} \bibnamefont{{Engelbracht}}},
et~al.,
\bibinfo{journal}{\apjs}
  \textbf{\bibinfo{volume}{154}}, \bibinfo{pages}{25} (\bibinfo{year}{2004}).

\bibitem[{\citenamefont{{Cutri} et~al.}(2003)\citenamefont{{Cutri},
  {Skrutskie}, {van Dyk}, {Beichman}, {Carpenter}, {Chester}, {Cambresy},
  {Evans}, {Fowler}, {Gizis} et~al.}}]{2MASS}
\bibinfo{author}{\bibfnamefont{R.~M.} \bibnamefont{{Cutri}}},
  \bibinfo{author}{\bibfnamefont{M.~F.} \bibnamefont{{Skrutskie}}},
  \bibinfo{author}{\bibfnamefont{S.}~\bibnamefont{{van Dyk}}},
et~al.,
\emph{\bibinfo{title}{{2MASS All Sky Catalog of point
  sources.}}} (\bibinfo{year}{2003}).

\bibitem[{\citenamefont{{Hartman} et~al.}(2006)\citenamefont{{Hartman},
  {Bersier}, {Stanek}, {Beaulieu}, {Kaluzny}, {Marquette}, {Stetson}, and
  {Schwarzenberg-Czerny}}}]{Hartman2006M33Variables}
\bibinfo{author}{\bibfnamefont{J.~D.} \bibnamefont{{Hartman}}},
  \bibinfo{author}{\bibfnamefont{D.}~\bibnamefont{{Bersier}}},
  \bibinfo{author}{\bibfnamefont{K.~Z.} \bibnamefont{{Stanek}}},
et~al.,
  \bibinfo{journal}{\mnras} \textbf{\bibinfo{volume}{371}},
  \bibinfo{pages}{1405} (\bibinfo{year}{2006}).

\bibitem[{\citenamefont{{Calzetti} et~al.}(1995)\citenamefont{{Calzetti},
  {Kinney}, {Ford}, {Doggett}, and {Long}}}]{Calzetti1995}
\bibinfo{author}{\bibfnamefont{D.}~\bibnamefont{{Calzetti}}},
  \bibinfo{author}{\bibfnamefont{A.~L.} \bibnamefont{{Kinney}}},
  \bibinfo{author}{\bibfnamefont{H.}~\bibnamefont{{Ford}}},
et~al.,
\bibinfo{journal}{\aj} \textbf{\bibinfo{volume}{110}},
  \bibinfo{pages}{2739} (\bibinfo{year}{1995}).

\bibitem[{\citenamefont{{Szeifert} et~al.}(1994)\citenamefont{{Szeifert},
  {Stahl}, {Wolf}, {Zickgraf}, and {Humphreys}}}]{Szeifert1994LBVsinM31andM33}
\bibinfo{author}{\bibfnamefont{T.}~\bibnamefont{{Szeifert}}},
  \bibinfo{author}{\bibfnamefont{O.}~\bibnamefont{{Stahl}}},
  \bibinfo{author}{\bibfnamefont{B.}~\bibnamefont{{Wolf}}},
et~al.,
\emph{\bibinfo{booktitle}{Astronomische  Gesellschaft Abstract
Series}}
\textbf{\bibinfo{volume}{10}}, \bibinfo{pages}{36} (1994).

\bibitem[{\citenamefont{{Humphreys} et~al.}(1987)\citenamefont{{Humphreys},
  {Jones}, and {Gehrz}}}]{Humphreys1987}
\bibinfo{author}{\bibfnamefont{R.~M.} \bibnamefont{{Humphreys}}},
  \bibinfo{author}{\bibfnamefont{T.~J.} \bibnamefont{{Jones}}},
  \bibnamefont{and} \bibinfo{author}{\bibfnamefont{R.~D.}
  \bibnamefont{{Gehrz}}}, \bibinfo{journal}{\aj} \textbf{\bibinfo{volume}{94}},
  \bibinfo{pages}{315} (\bibinfo{year}{1987}).

\bibitem[{\citenamefont{{Humphreys} et~al.}(2006)\citenamefont{{Humphreys},
  {Jones}, {Polomski}, {Koppelman}, {Helton}, {McQuinn}, {Gehrz}, {Woodward},
  {Wagner}, {Gordon} et~al.}}]{Humphreys2006}
\bibinfo{author}{\bibfnamefont{R.~M.} \bibnamefont{{Humphreys}}},
  \bibinfo{author}{\bibfnamefont{T.~J.} \bibnamefont{{Jones}}},
  \bibinfo{author}{\bibfnamefont{E.}~\bibnamefont{{Polomski}}},
et~al.,
\bibinfo{journal}{\aj} \textbf{\bibinfo{volume}{131}},
  \bibinfo{pages}{2105} (\bibinfo{year}{2006}).

\bibitem[{\citenamefont{{Osterbrock} and
  {Ferland}}(2006)}]{OsterbrockFerland2006}
\bibinfo{author}{\bibfnamefont{D.~E.} \bibnamefont{{Osterbrock}}}
  \bibnamefont{and} \bibinfo{author}{\bibfnamefont{G.~J.}
  \bibnamefont{{Ferland}}}, \emph{\bibinfo{title}{{Astrophysics of gaseous
  nebulae and active galactic nuclei}}}
(Univ. Sci. Books, Sausalite, \bibinfo{year}{2006}).

\bibitem[{\citenamefont{{O'Donnell}}(1994)}]{ODonnell1994ReddenCurve}
\bibinfo{author}{\bibfnamefont{J.~E.} \bibnamefont{{O'Donnell}}},
  \bibinfo{journal}{\apj} \textbf{\bibinfo{volume}{422}}, \bibinfo{pages}{158}
  (\bibinfo{year}{1994}).

\bibitem[{\citenamefont{{Chentsov} and
  {Sarkisyan}}(2007)}]{Chentsov2007SpectralAtlas}
\bibinfo{author}{\bibfnamefont{E.~L.} \bibnamefont{{Chentsov}}}
  \bibnamefont{and} \bibinfo{author}{\bibfnamefont{A.~N.}
  \bibnamefont{{Sarkisyan}}}, \bibinfo{journal}{Astrophysical Bulletin}
  \textbf{\bibinfo{volume}{62}}, \bibinfo{pages}{257} (\bibinfo{year}{2007}).

\bibitem[{\citenamefont{{Didelon}}(1982)}]{Didelon1982BstarsEW}
\bibinfo{author}{\bibfnamefont{P.}~\bibnamefont{{Didelon}}},
  \bibinfo{journal}{\aaps} \textbf{\bibinfo{volume}{50}}, \bibinfo{pages}{199}
  (\bibinfo{year}{1982}).

\bibitem[{\citenamefont{{Flower}}(1996)}]{Flower1996BolometricCorrection}
\bibinfo{author}{\bibfnamefont{P.~J.} \bibnamefont{{Flower}}},
  \bibinfo{journal}{\apj} \textbf{\bibinfo{volume}{469}}, \bibinfo{pages}{355}
  (\bibinfo{year}{1996}).

\bibitem[{\citenamefont{{Claret}}(2006)}]{Claret2006Evolution0.007}
\bibinfo{author}{\bibfnamefont{A.}~\bibnamefont{{Claret}}},
  \bibinfo{journal}{\aap} \textbf{\bibinfo{volume}{453}}, \bibinfo{pages}{769}
  (\bibinfo{year}{2006}).

\bibitem[{\citenamefont{{Groh} et~al.}(2009)\citenamefont{{Groh}, {Hillier},
  {Damineli}, {Whitelock}, {Marang}, and {Rossi}}}]{Groh2009AGCarNature}
\bibinfo{author}{\bibfnamefont{J.~H.} \bibnamefont{{Groh}}},
  \bibinfo{author}{\bibfnamefont{D.~J.} \bibnamefont{{Hillier}}},
  \bibinfo{author}{\bibfnamefont{A.}~\bibnamefont{{Damineli}}},
et~al.,
  \bibinfo{journal}{\apj} \textbf{\bibinfo{volume}{698}}, \bibinfo{pages}{1698}
  (\bibinfo{year}{2009}).

\end{thebibliography}
\end{document}